\documentclass[structabstract]{aa}  
\usepackage{graphicx}
\usepackage{txfonts}
\usepackage{hyperref}
\usepackage{longtable,lscape, xcolor}
\begin{document}
   \title{How are Forbush decreases related to interplanetary magnetic field enhancements ?}


  \author{K. P. Arunbabu\inst{1,2}, H. M. Antia\inst{2,3}, S. R. Dugad\inst{2,3}, S. K. Gupta\inst{2,3}, Y. Hayashi\inst{2,4}, S. Kawakami\inst{2,4}, P. K. Mohanty\inst{2,3}, A. Oshima\inst{2,5}, P. Subramanian\inst{1,2} 
           }

   \institute{Indian Institute of Science Education and Research, Dr. Homi Bhabha Road, Pashan, Pune 411 021, India \and
The GRAPES--3 Experiment, Cosmic Ray Laboratory, Raj Bhavan, Ooty 643 001, India \and 
Tata Institute of Fundamental Research, Homi Bhabha Road, Mumbai 400 005, India \and
Graduate School of Science, Osaka City University, Osaka 558-8585, Japan           \and 
College of Engineering, Chubu University, Kasugai, Aichi 487-8501, Japan
}

\authorrunning{Arunbabu et al.}
\titlerunning{Forbush decreases related to IMF}
   \date{}

 
  \abstract
   {}
    {{{ A Forbush decrease (FD) is a transient decrease followed by a gradual recovery  in the observed galactic cosmic ray intensity.}} We seek to understand the relationship between the FDs      and near-Earth interplanetary magnetic field (IMF) enhancements associated
     with solar coronal mass ejections (CMEs).}
   {We used muon data at cutoff rigidities ranging from 14 to 24 GV from the
    GRAPES-3 tracking muon telescope to identify FD events. We selected those
    FD events that have a reasonably clean profile, and magnitude $>$\,0.25\%.
    We used IMF data from ACE/WIND spacecrafts. We looked for correlations
    between the FD profile and that of the one-hour averaged IMF. We wanted to find out whether if
    the diffusion of high-energy protons into the large scale magnetic field
    is the cause of the lag observed between the FD and the IMF.}
   {The enhancement of the IMF associated with FDs occurs mainly in the
    shock-sheath region, and the turbulence level in the magnetic field is
    also enhanced in this region. The observed FD profiles look remarkably
    similar to the IMF enhancement profiles. The FDs typically lag behind the IMF
    enhancement by a few hours. The lag corresponds to the time taken by
    high-energy protons to diffuse into the magnetic field enhancement via
    cross-field diffusion.}
    {Our findings show that high-rigidity FDs associated with CMEs are
     caused primarily by the cumulative diffusion of protons across the
     magnetic field enhancement in the turbulent sheath region between
     the shock and the CME.}

    \keywords{cosmic rays, Forbush decrease, interplanetary magnetic field,
              coronal mass ejection}

   \maketitle
%

\section{Introduction}

Forbush decreases (FDs), are short-term decreases in the intensity of
galactic cosmic rays that were first observed by Forbush (1937, 1938).
It was the work of Simpson using neutron monitors (Simpson, 1954) that showed that the origin of the FDs was in the interplanetary (IP)
medium. Solar transients such as the coronal mass ejections (CMEs) cause enhancements in the interplanetary magnetic field (IMF).
The near-Earth manifestation of a CME from the Sun typically has two
major components: i) the interplanetary counterpart of CME (commonly
called an ICME), and ii) the shock, which is driven ahead of it. Both
the shock and the ICME cause significant enhancement in the IMF. Interplanetary CMEs,
which possess some well defined criteria such as reduction in plasma
temperature and smooth rotation of magnetic field are called magnetic
clouds (e.g., Burlaga et al. 1981; Bothmer \& Schwenn 1998).

Correlations between the parameters characterizing FDs and solar wind
parameters have been a subject of considerable study. Belov et al.
(2001) and Kane (2010) maintain that there is a reasonable correlation
between the FD magnitude and the product of maximum magnetic field and
maximum solar wind velocity. Dumbovi\'{c} et al. (2012) also found
reasonable correlation between the FD magnitude $|\rm FD|$, and duration
with the solar wind parameters such as the amplitude of magnetic field
enhancement B, amplitude of the magnetic field fluctuations $\rm \delta B$,
maximum solar wind speed associated with the disturbance v, and duration of
the disturbance $\rm t_B$. We note that the FD magnitude also depends
strongly on other solar wind parameters like the velocity of the CME,
turbulence level in the magnetic field, size of the CME, etc. The
contributions of these parameters are explained in the CME-only cumulative
diffusion model described in Arunbabu et al. (2013).

Arunbabu et al. (2013) described the CME-only cumulative diffusion
model for FDs, where the cumulative effects of diffusion of cosmic
ray protons through the turbulent sheath region as the CME propagated
from the Sun to the Earth was invoked to explain the FD magnitude.
However, the diffusion was envisaged to occur across an idealized thin
boundary. In this paper we relax the ideal, thin boundary assumption and
examine the detailed relationship between the FD profile and the IMF
compression. 

\section{Data analysis}\label{DA}
\subsection{The GRAPES-3 experiment}\label{G3}
The GRAPES-3 experiment is located at Ooty ($11.4^{\circ}$N latitude,
$76.7^{\circ}$E longitude, and 2200\,m altitude) in India. It contains
two major components, first an air shower array of 400 scintillation
detectors (each 1\,m$^2$), with a distance of 8\,m between adjacent
detectors deployed in a hexagonal geometry (Gupta et al. 2005, Mohanty
et al. 2009, Mohanty et al. 2012). The GRAPES-3 array is designed to
measure the energy spectrum and composition of the primary cosmic rays
in the energy region from 10\,TeV to 100\,PeV (Gupta et al. 2009,
Tanaka et al. 2012). The second component of the GRAPES-3 experiment, a
large area tracking muon telescope is a unique instrument used to search
for high-energy protons emitted during the active phase of a solar flare
or a CME. The muon telescope provides a high statistics, directional
measurement of the muon flux. The GRAPES-3 muon telescope covers an
area of 560\,m$^2$, consisting of a total of 16 modules, each 35\,m$^2$
in area. The energy threshold of the telescope is sec\,($\theta$)\,GeV
for the muons arriving along a direction with zenith angle $\theta$.
The observed muon rate of $\sim3000$\,s$^{-1}$ per module yields a total
muon rate of $\sim3\times10^6$ min$^{-1}$ for the entire telescope (Hayashi
et al. 2005, Nonaka et al. 2006). This high rate permits even a small
change of $\lesssim0.1\%$ in the muon flux to be accurately measured over
a time scale of $\sim5$\,min, after appropriate corrections are applied
for the time dependent variation in the atmospheric pressure (Mohanty et
al. 2013).

We identified the FDs using the data from the GRAPES-3 muon telescope.
By using the tracking capability of the muon telescope the direction of
detected muons are binned into nine different solid angle directions, named
NW (northwest), N (north), NE (northeast), W (west), V (vertical), E (east), SW (southwest), S (south), and SE (southeast). The cutoff rigidity due to
the geomagnetic field at Ooty is 17\,GV along the vertical direction and
varies from 14 to 42\,GV across the 2.2\,sr. field of view  of the muon
telescope. Details of the muon telescope are given in Hayashi et al. (2005),
Nonaka et al. (2006), and Subramanian et al. (2009). 

\subsection{Broad shortlisting criteria}\label{SL1}
{ The GRAPES-3 muon telescope has observed a large number of FD events exhibiting a variety of characteristics. We now describe the broad criteria that we use to shortlist events used for analysis in this paper. In addition to the criteria described here, we will have occasion to apply further criteria, which will be described in subsequent sections. We have examined all FD
events observed by the GRAPES-3 muon telescope during the years 2001 --
2004. We shortlist events that have a clean FD profile and FD magnitude
$>$\,0.25\%, and are also associated with an enhancement in the near-Earth interplanetary magnetic field. Here, the term ``clean profile" is used to refer to an FD event
characterized by a sudden decrease and a gradual recovery in the cosmic
ray flux. Although 0.25 \% might seem like a small number, according to Arunbabu et al. (2013) these are fairly significant events in the GRAPES-3 data, given its high sensitivity. This yields a sample of 65 events, which is used in the analysis reported in \S~\ref{Corrpeak}. We note that the event of 29 September 2001 is not included in this list 
since it was associated with many IMF enhancements, which could be due to
multiple Halo and partial halo CMEs. Similarly, the event of 29 October 2003 is not included (even though it was the biggest FD event observed in the solar cycle 23) since the near-Earth magnetic field data is incomplete for this event.}

\begin{figure}
\centering

\includegraphics[width = 0.95\columnwidth]{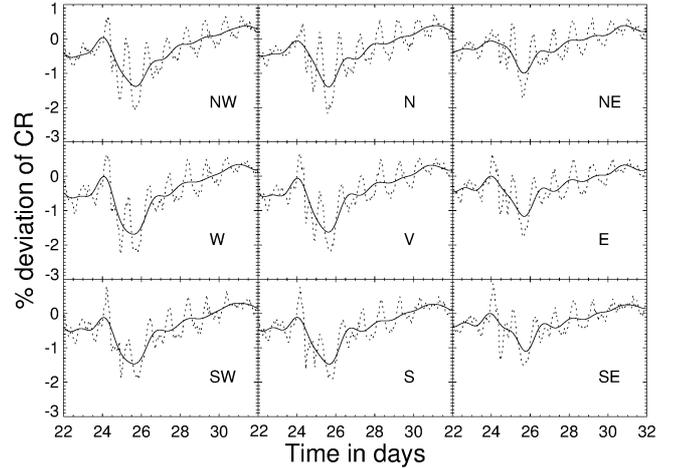}
\caption{FD event on 24 Nov 2001. Percent deviation in the muon
         flux for different directions are shown in separate
         panels. The solid line shows percent deviation for
         filtered data and the dotted line shows the same for
         unfiltered data. }\label{unf}
\end{figure}

We used GRAPES-3 data summed over a time interval of one hour for each
direction. This improves the signal-to-noise ratio, but the diurnal
variation in the muon flux is still present. We used a low-pass filter
to remove oscillations having frequency $>$\,1\,d$^{-1}$. This filter
was explained in Subramanian et al. (2009). The measured variation
of the muon flux in percent for the 24 November 2001 event is shown in
Figure \ref{unf}, where the dotted black lines are the unfiltered data and
the solid black lines are the filtered data after using the low-pass filter
to remove the frequencies $>$\,1\,d$^{-1}$. Although the filtering may
change the amplitude of the decrease and possibly shift the onset time
for the FD by a few hours in some cases, it is often difficult to
determine whether these differences are artifacts of filtering or if the
unfiltered data showed different amplitude because a diurnal oscillation
happened to have the right phase so as to enhance or suppress the
amplitude of the FD. Some fluctuations in the muon flux could be due to
FD and associated events, but it is unlikely that they will be periodic
in nature and are not likely to be affected by the filter. 

\begin{figure*}
\sidecaption
\includegraphics[width=12cm]{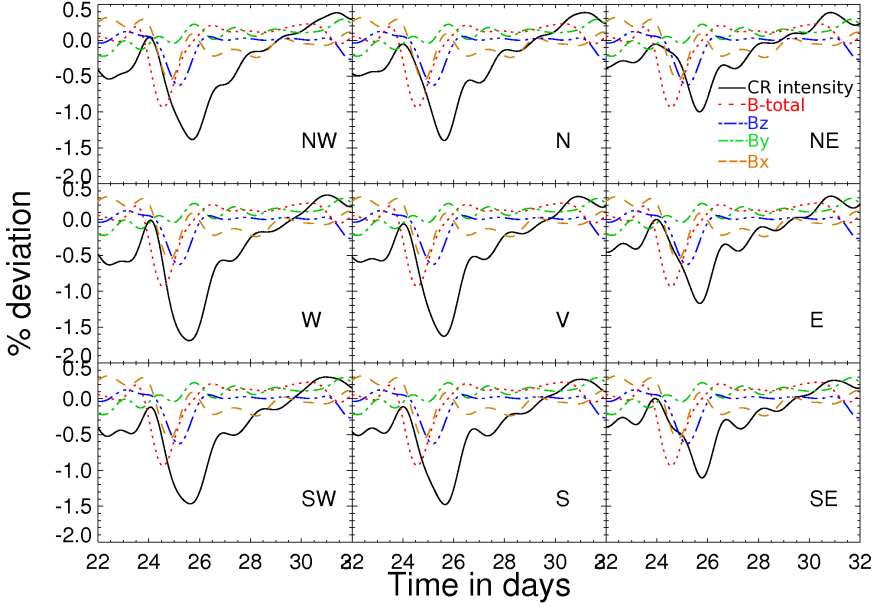} 
\caption{The FD event of { 24 Nov 2001 } and the magnetic field for
         9 directions in GRAPES-3 muon telescope. Black solid line
         is percentage deviation of cosmic ray intensity in each
         direction. The red-dotted, blue-dash-dot-dotted, green-dash-dotted
         and orange-dash lines are percentage deviation of IMFs $\rm
         B_{total}$, $\rm B_z$,  $\rm B_y $ and  $\rm B_x$ respectively,
         that are scaled down by a factor of 10 to fit in the frame.  }
\label{FDi}
\end{figure*}

The FDs that we study are associated with near-Earth CME counterparts,
which contribute to significant increases in the IMFs. We intend to
investigate the relation between these IMF enhancements and FDs. We
used the IMF data observed by the ACE and WIND spacecraft available
from the \href{http://omniweb.gsfc.nasa.gov/}{ \textit{\rm OMNI}}
database. We used hourly resolution data on magnetic field
$\rm B_{total}$, $\rm B_x$, $\rm B_y$, $\rm B_z$ in the geocentric
solar ecliptic (GSE) coordinate system: $\rm B_{total}$ is the
magnitude of the magnetic field, $\rm B_x$ is the magnetic field
component along the Sun-Earth line in the ecliptic plane pointing
towards the Sun, $\rm B_z$ the component parallel to the ecliptic
north pole, and $\rm B_y$ the component in the ecliptic plane pointing
towards dusk. For a consistent data analysis we applied the same
low-pass filter to the magnetic field data as we did to the muon flux,
which removes any oscillations having frequency $>$ 1\,d$^{-1}$. Since
FD events are associated with enhancements in the IMF, we use the
quantity $\rm 100-|B|$ and calculate the average value and the percent
deviation of this quantity over the same data interval as the FD. This
effectively flips the magnetic field increase and makes it appear
as a decrease, enabling easy comparison with the FD profile. { Figure
\ref{FDi} shows the FD event on 24 November 2001, \ref{FDiap11} shows the FD event on 11 April 2001, and \ref{FDimy23} shows the FD event on 23 May 2002,} together with the IMF
data processed in this manner.  \\

\section{Correlation of FD magnitude with peak IMF}\label{Corrpeak} 

Before studying the detailed relationship between the IMF and FD profiles,
we examine the relationship between the peak IMF and the FD magnitudes.
{ For this study we considered all 65 FD events shortlisted using the criteria explained in section \ref{SL1}.}  The FD magnitude for a given direction is calculated as the
difference between the pre-event intensity of the cosmic rays and the
intensity at the minimum of the decrease. We examine the corresponding
IMF during these events. We denote $\rm B_y$ and $\rm B_z$ ``perpendicular''
fields, because they are tangential to a flux rope CME approaching the
Earth. They are perpendicular to a typical cosmic ray proton that seeks
to enter the CME radially; it will therefore have to cross these
perpendicular fields. We study the relation between the FD magnitude and
the peak of the total magnetic field $\rm B_{total} = \,(B_x^2 + B_y^2 +
B_z^2)^{1/2}$ and the peak of the net perpendicular magnetic field $\rm
B_p \, = \,(B_y^2 + B_z^2)^{1/2}$.

\begin{figure}[h]
\centering
\includegraphics[width =0.95\columnwidth]{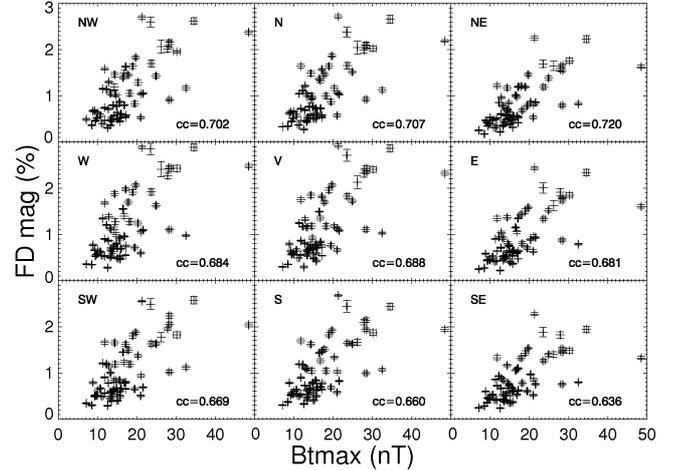}
\caption{ Correlation of maximum total magnetic field in the magnetic
         field enhancement to FD magnitude observed in different directions using GRAPES-3.}\label{corBT}
\end{figure}

\begin{figure}[h]
\centering
\includegraphics[width =0.95\columnwidth]{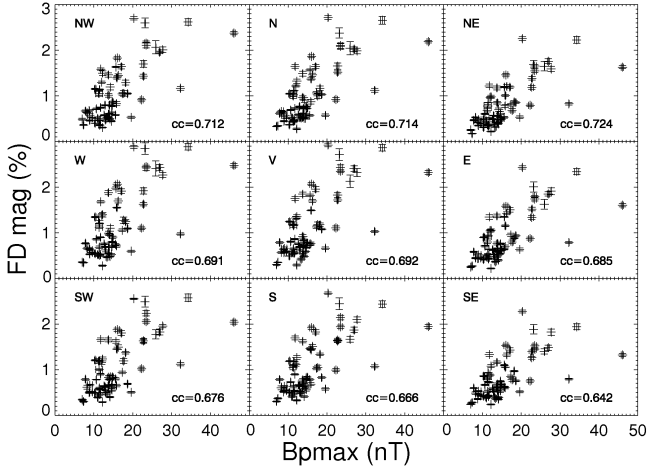}
\caption{ Correlation of maximum perpendicular magnetic field in the
         magnetic field enhancement to FD magnitude observed in different directions using GRAPES-3.}\label{corBP}
\end{figure}

The correlation coefficients of the peak $\rm B_{total}$ with the FD
magnitude for different directions are listed in Table\,\ref{corrT} and
shown in Figure \ref{corBT}. The correlation coefficients of the peak
$\rm B_p$ with FD magnitude are listed in Table\,\ref{corrT} and shown
in Figure \ref{corBP}. We find that the correlation coefficient between
peak $\rm B_p$ and peak $\rm B_{total}$ with the FD magnitude ranges from
63\% to 72\%.  { We note that the correlations in Figures 3 and 4 are fairly similar because the longitudinal magnetic field ($B_{x}$) is fairly small for most events. We will have further occasion to discuss this in \S~\ref{crossfield}.} { We also carried out the same study using Tibet neutron monitor data; this yields a correlation coefficient of 60.0\% and 61.9\% respectively for $\rm B_{total}$ and $\rm B_{p}$. The error on the correlation coefficients are calculated using Eq \ref{er} below, and are listed in Table~\ref{corrT}:

\begin{equation}
err = \sqrt{\frac{1-(cc)^2}{n-2}}\, . \label{er}
\end{equation} 

Here  cc is the correlation coefficient, n-2 gives the degree of freedom, and n is the number of points considered for the correlation.}

From the CME-only cumulative diffusion model described in
Arunbabu et al. (2013) we know that the FD magnitude depends on various
parameters associated with CME, such as velocity of CME, turbulence level
in the magnetic field, and the size of the CME. It is thus not surprising
that the FD magnitude correlates only moderately with the peak value of
the IMF.

\begin{table}[h]
\centering
\caption{Correlation of the FD magnitude with the maximum total and
         perpendicular IMF. For each direction of GRAPES-3, the
         correlation is calculated along with the standard error (err) in the next column. {  The last row shows the correlation coefficient calculated using data from the Tibet neutron monitor.}}\label{corrT}
{  \begin{tabular}{l c c c c c} \hline \hline
Direction &Cut-off & \multicolumn{2}{c}{ $\rm B_{total}$  } & \multicolumn{2}{c}{  $\rm B_p$ }\\
 &  Rigidity &  \multicolumn{2}{c}{ Correlation } &  \multicolumn{2}{c}{ Correlation } \\
 &   (GV) &  coeff. & err &  coeff. & err  \\ \hline 
NW & 15.5 & 0.702 & 0.090 & 0.712 & 0.089\\
N  & 18.7 & 0.707 & 0.090 & 0.714 & 0.089\\  
NE & 24.0 & 0.720 & 0.088 & 0.724 & 0.088\\
W  & 14.3 & 0.684 & 0.093 & 0.691 & 0.092\\
V  & 17.2 & 0.688 & 0.092 & 0.692 & 0.092\\
E  & 22.4 & 0.681 & 0.093 & 0.685 & 0.093\\ 
SW & 14.4 & 0.669 & 0.094 & 0.676 & 0.094\\
S  & 17.6 & 0.660 & 0.095 & 0.666 & 0.095\\ 
SE & 22.4 & 0.636 & 0.098 & 0.642 & 0.097\\ \hline \hline
Tibet & 14.1 & 0.600 & 0.102&  0.619 & 0.100\\ \hline \hline
\end{tabular} }
\end{table}

\section{IMF compression: shock-sheath or ICME?} \label{ipmc}

\begin{figure*}
\sidecaption
\includegraphics[width=12cm]{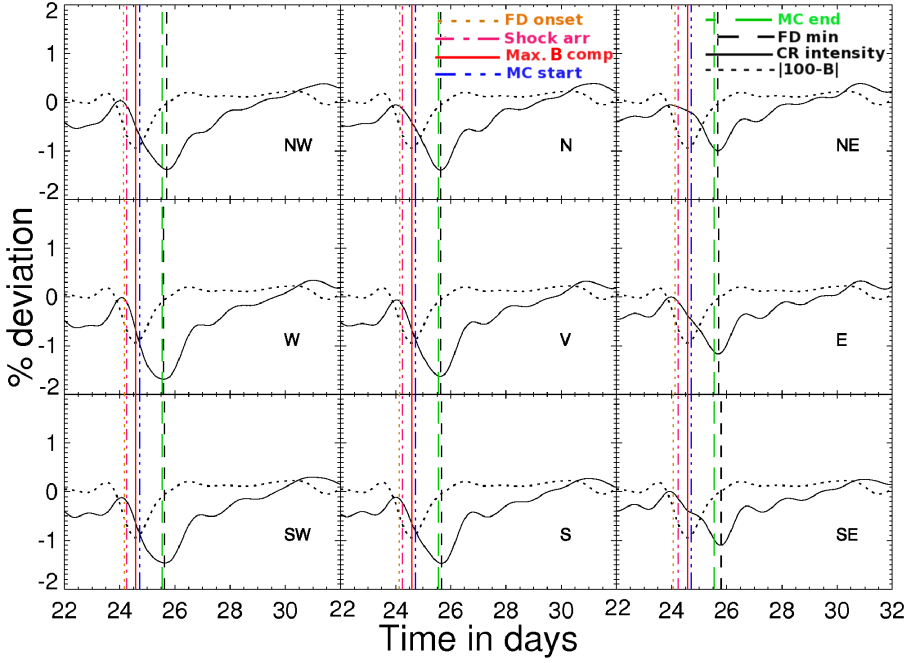} 
\caption{FD event on 24 November 2001. The black solid line denotes
         percentage deviation of the cosmic ray intensity, black
         dotted line percentage deviation of total magnetic field
         $\rm |100 - B|$ as explained in \S\ref{DA} which is
         scaled down to fit the frame. The vertical brown-dotted,
         magenta-dash-dotted, blue-dash-dot-dotted, green-long-dashed,
         and black-dashed lines denote the timings corresponding to
         the FD onset, shock arrival, magnetic cloud start, magnetic
         cloud end, and FD minimum, respectively. The solid red
         vertical line corresponds to the maximum of the magnetic
         field compression.  }
\label{timi}
\end{figure*}

As mentioned earlier, we considered FD events associated with the
magnetic field enhancements that are due to the shock propagating
ahead of the ICME. {  An example of this is shown in Figures \ref{timi}, \ref{timiapr11}, and  \ref{timimay23}}
where the nine different panels show the cosmic ray flux (FD profile)
of the nine directions of the GRAPES-3 muon telescope for the FD observed on
24 November 2001. It is clear that the magnetic field compression responsible for the FD is in
the sheath region, i.e., the region between the shock and magnetic cloud.

\begin{figure}[h]
\centering
\includegraphics[width =0.95\columnwidth]{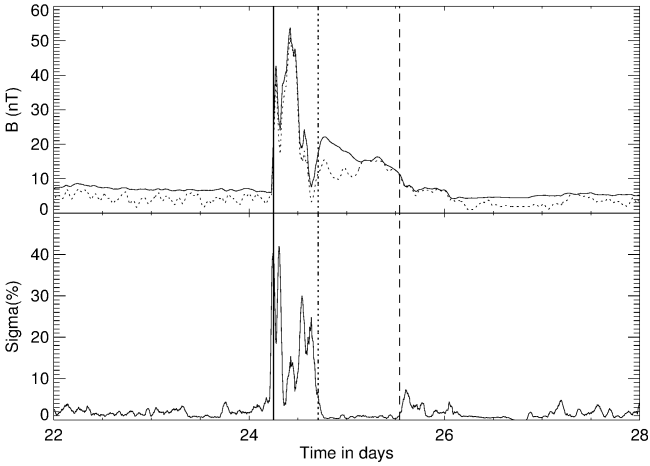}
\caption{\bf Magnetic field compression associated with the FD event on 24
         November 2001. In the first panel the continuous line denotes $\rm
         B_{total}$ and the dotted line denotes $\rm B_p$. The continuous
         line in the second panel shows the turbulence level for $\rm
         B_{total}$. In both panels the vertical solid, dotted, and dashed lines
         corresponds to shock arrival time, magnetic cloud start time, magnetic cloud end time, respectively. \label{tur}}
\end{figure}

The CME-only model described in Arunbabu et al. (2013) deals with the
diffusion of cosmic rays through the turbulent magnetic field in the
sheath region. The cross-field diffusion coefficient depends on the
rigidity of the proton and the turbulence level in the magnetic field (e.g., Candia \& Roulet 2004). The turbulence level in the magnetic
field is an important parameter in this context. We have calculated
the turbulence level using one-minute averaged data from the ACE/WIND
spacecraft available from the \href{http://omniweb.gsfc.nasa.gov/}{
\textit{\rm OMNI}} data base. To calculate the turbulence level $\sigma$
we use a one-hour running average of the magnetic field ($\rm B_0$) and
the fluctuation of the IMF around this average ($\rm B_{tur} = B -B_0$).
We define the quantity $\sigma$ as
\begin{equation}
\rm \sigma \, = \,\left( \frac {\langle B_{tur}^2 \rangle}{B_0^2}\right) ^{0.5}, \label{dysig}
\end{equation}

where $\rm \langle B_{tur}^2 \rangle$ denotes the average of $\rm
B_{tur}^2$ over the one-hour window. Figure \ref{tur} shows a
representative event. The top panel shows the one-minute average
magnetic field for 21--30 November 2001. The bottom panel in this
figure shows the turbulence level $\sigma$ calculated for this
event. We note that the magnetic
field compression responsible for the FD occurs in the shock
sheath region, i.e., the region between the shock and the magnetic
cloud. The turbulence level enhancement also occurs in this region.

{  For events selected using the criteria described in \S\,2, we narrow down the ones that have a  well-defined shock and associated magnetic cloud. Such events allow us to clearly distinguish the shock, sheath, and CME regions associated with the magnetic field compression. We studied ten such FD events, which are listed in Table \ref{tabtim}. } The
timings of the shock, maximum of the magnetic field compression,
magnetic cloud start and end timings, along with the FD onset times
for different directions are given in Table\,\ref{tabtim}. The peak of
the magnetic field enhancement in the filtered data generally occurs
before the start of the magnetic cloud or at the start of the magnetic
cloud, whereas in the unfiltered data the enhancement lies in the
sheath region. We note that the filtering procedure using the low-pass
filter shifts the maximum by a small amount (-5 to 10 hours).

\begin{table*}
\centering
\caption{  Shock arrival time, time of maximum magnetic field enhancement,
         magnetic cloud start and end timings and FD onset timings for different directions for FD events that have a well-defined shock and magnetic clouds associated with them. }\label{tabtim}
\begin{tabular}{lccccccccccccc} \hline \hline
Event & \multicolumn{9}{c} {FD onset} & Shock &Maximum of  & MC & MC \\   
      & NW & N & NE & W & V & E & SW & S & SE & arrival &Mag. compre. & Start & end \\ \hline 
2001 Apr 04  & 04.30  & 04.33 & 04.29 & 04.3 &04.34 &04.32 &04.37 & 04.34 &04.25 & 04.61 & 04.79 & 04.87 & 05.35 \\
2001 Apr 11  & 11.54 & 11.58 & 11.67 & 11.47 &11.50 &11.54 &11.35 & 11.43 &11.52 & 11.58 & 12.00 &  11.958 & 12.75 \\
2001 Aug 17  & 17.18 & 17.08 & 17.05 & 16.97 &16.94 &16.92 &17.00 & 16.98 &16.97 & 17.45 & 17.87 & 18.00 & 18.896 \\
2001 Nov 24  & 24.13 & 24.14 & 24.13 & 24.17 &24.14 &24.12 &24.17 & 24.13 &24.06 & 24.25 & 24.58 &  24.708 & 25.541 \\
2002 May 23  & 23.13 & 23.08 & 23.00 & 23.17 &23.09 &23.04 &23.21 & 23.13 &23.09 & 23.44 & 23.58 &  23.896 & 25.75 \\
2002 Sep 07  & 07.71 & 07.72 & 07.67 & 07.62 &07.62 &07.63 &07.66 & 07.68 &07.71 & 07.6  & 08.00 & 07.708 & 08.6875 \\
2002 Sep 30  & 30.56 & 30.45 & 30.43 & 30.56 &30.48 &30.42 &30.52 & 30.45 &30.36 & 30.31 & 31.16 & 30.917 & 31.6875 \\
2003 Nov 20  & 19.89 & 21.34 & 21.12 & 20.08 &20.45 &20.86 &20.29 & 20.43 &20.63 & 20.31 & 20.67 & 21.26 & 22.29 \\
2004 Jan 21  & 22.09 & 22.03 & 21.96 & 21.98 &22.00 &21.97 &21.83 & 21.91 &21.90 & 22.09 & 22.50 &  22.58 & 23.58 \\
2004 Jul 26  & 26.60 & 26.64 & 26.75 & 26.59 &26.65 &26.75 &26.67 & 26.77 &26.86 & 26.93 & 27.29 & 27.08 & 28.00 \\  \hline \hline
\end{tabular} 
\end{table*}

It is clear from {  Figures\,\ref{timi}, \, \ref{timiapr11}, \, \ref{timimay23}, \,\ref{tur}, and Table\,\ref{tabtim} }
that the peak of the magnetic field compression responsible for the FD
lies in the sheath region, and the turbulence level is also enhanced in
this region. This is in broad agreement with Richardson \& Cane (2011).

\section{How similar are the FD and the IMF profiles?}
\begin{figure} [h]
\centering
\includegraphics[width =0.95\columnwidth]{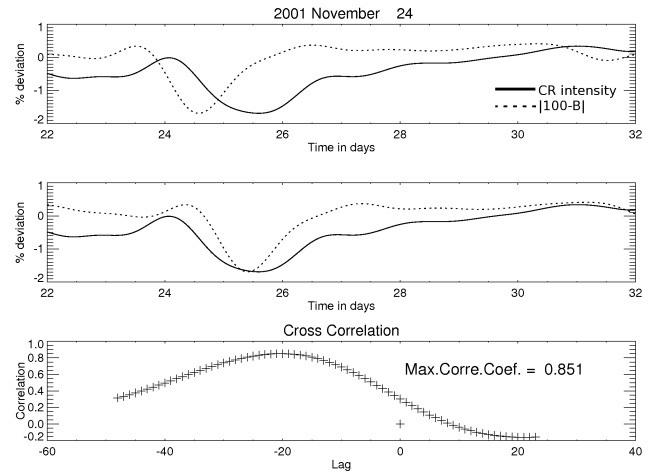}
\caption{Cross correlation of the cosmic ray flux with $\rm B_{total}$.
         The top panel shows the percentage deviation of cosmic ray flux
         using solid black lines and the magnetic field using dotted
         black lines (scaled to fit in the frame). The middle panel
         shows the same with magnetic field shifted to the right
         corresponding to the time lag and the bottom panel shows the
         correlation coefficient for different lags.}\label{24nov} 
\end{figure}

One of the near-Earth effects of a CME is the compression of (and
consequent increase in) the IMF. The IMF measured by spacecrafts such
as WIND and ACE can detect these magnetic field compressions. We
investigate the relation of these magnetic field compressions to
the FD profile. We work with the hourly resolution IMF data from
the ACE and WIND spacecrafts obtained from the
\href{http://omniweb.gsfc.nasa.gov/}{ \textit{OMNI}} database.
Applying the low-pass filter described in \S\ref{DA} to this data
yields a combined magnetic field compression comprising the shock
and ICME/magnetic cloud. A visual comparison of the FD profile with
the magnetic field compression often reveals remarkable similarities.
To quantify the similarity between these two profiles, we studied
the cross correlation of the cosmic ray intensity profile with the
IMF profile. 
In order to do this, we shift the magnetic field profile
(with respect to the FD profile) by amounts ranging from -36 hours
to 12 hours. We identify the peak correlation value and the shift
corresponding to this value is considered to be the time lag between
the IMF and the cosmic ray FD profile. Most of the FD events exhibit
correlations $\geq$\,60\% with at least one of the four IMF
components ($\rm B_{total}$, $\rm B_x$, $\rm B_y$, $\rm B_z$).
{  Examples of the correlation between the FD profile
and $\rm B_{total}$ for the events of 24 November 2001, 11 April 2001, and 23 May 2002 are shown in Figures \ref{24nov}, \ref{11apr},  and \ref{23may}, respectively. The top panel in these figures shows
the percentage deviation of the cosmic ray flux and the $\rm B_{total}$.
The percentage deviation of $\rm B_{total}$ is scaled to fit in the frame.
The middle panel shows the same percentage deviations, but the magnetic
field is shifted by the peak correlation lag and the bottom panel shows
the correlation coefficients corresponding to different lags. The correlation lag means that the IMF profile 
precedes the FD profile by 21 hours for the 24 November 2001 event, by 19 hours for the 11 April 2001 event and 13 hours for the 23 May 2003 event.}  { In particular, Table \ref{T24} gives the maximum cross correlation values obtained for different FD profile of different directions of GRAPES-3 to the total magnetic field compression and different components of magnetic field for the 24 November 2001 event. The same quantities for Tibet neutron monitor data are also shown in last row of this table. In further
discussion we consider only those events showing a cross correlation $\geq$\,70\% for lags between -36 to 12 hours. The events thus shortlisted are presented in Table\,\ref{T1}. From these, we will further narrow down events for which the FD profile exhibits a high correlation with the perpendicular component of the IMF.}

\begin{table*}
\centering
\caption{\label{T24}  Correlation of FD profile in different direction observed in GRAPES-3 with magnetic fields for the 24 November 2001 event. The numbers in the last row are derived from the Tibet neutron monitor data.}
\begin{tabular}{lccccccccccccccc}\hline \hline
Instrument & dir. & Rg & FD mag. &  \multicolumn{12}{c}{correlation} \\  
& & (GV) & (\%) & \multicolumn{3}{c}{$B_{total}$} & \multicolumn{3}{c}{$B_{x}$}& \multicolumn{3}{c}{$B_{y}$}& \multicolumn{3}{c}{$B_{z}$}\\   
& &  &  & Corr. & err&  Lag & Corr.& err& Lag & Corr. & err& Lag & Corr. & err& Lag \\
& &  &  & (\%) & (\%)& (hrs)& (\%) & (\%)& (hrs)& (\%) & (\%)&  (hrs)& (\%) & (\%)& (hrs)\\\hline 
	&  NW & 15.5 & 2.60 &  80.4 & 3.8 &  -23 &    27.3  & 6.2 &  -13 &    36.3 & 6.0 &  -35 &    73.0 & 4.4 &  -13\\
	&  N  & 18.7 & 2.38 &  82.3 & 3.7 &  -23 &    29.2  & 6.2 &  -16 &    37.9 & 6.0 &  -35 &    73.8 & 4.4 &  -12\\
GRAPES-3&  NE & 24.0 & 1.69 &  80.7 & 3.8 &  -26 &    26.5  & 6.2 &  -19 &    39.3 & 5.9 &  -35 &    72.6 & 4.5 &  -14\\
	&  W  & 14.3 & 2.85 &  85.3 & 3.4 &  -21 &    30.1  & 6.1 &  -12 &    40.3 & 5.9 &  -30 &    75.7 & 4.2 &  -11\\
	&  V  & 17.2 & 2.71 &  85.3 & 3.4 &  -21 &    29.4  & 6.2 &  -14 &    40.5 & 5.9 &  -28 &    76.3 & 4.2 &  -11\\
	&  E  & 22.4 & 2.00 &  82.1 & 3.7 &  -24 &    29.7  & 6.2 &  -17 &    39.3 & 5.9 &  -22 &    75.5 & 4.2 &  -13\\
	&  SW & 14.4 & 2.49 &  84.6 & 3.4 &  -21 &    29.1  & 6.2 &  -12 &    41.0 & 5.9 &  -31 &    74.3 & 4.3 &  -11\\
	&  S  & 17.6 & 2.44 &  84.9 & 3.4 &  -21 &    31.0  & 6.1 &  -14 &    39.0 & 6.0 &  -27 &    76.6 & 4.1 &  -11\\
	&  SE & 22.4 & 1.89 &  81.1 & 3.8 &  -26 &    32.4  & 6.1 &  -20 &    36.8 & 6.0 &  -22 &    77.1 & 4.1 &  -14\\ \hline
Tibet	&     & 14.1 & 4.28 &  87.9 & 3.1 &  -20 &    27.7  & 6.2 &  -11 &    38.9 & 6.0 &  -20 &    75.6 & 4.2 &  -8 \\ \hline \hline
\end{tabular}
\end{table*}

\subsection{Cross-field diffusion into the ICME through the sheath}\label{crossfield}

The time lag between the cosmic ray flux and the IMF occurs because the high-energy protons do not respond to magnetic field
compressions immediately; they are subjected to the classical magnetic
mirror effect arising from the gradient in the longitudinal magnetic
field and to turbulent cross-field (also referred as perpendicular)
diffusion (e.g., Kubo \& Shimazu, 2010). Here, we concentrate only on
the cross-field diffusion of the high-energy protons through the
turbulent sheath region between the shock and the CME. As discussed
earlier, we have identified the IMF compression to be comprised mainly
of this sheath region; we therefore use the observed values of the
mean field and turbulent fluctuations in the sheath region to
calculate representative diffusion timescales for cosmic rays. The
time delay between the IMF compression and the FD profile (i.e., the
correlation lag) can be interpreted as the time taken by the cosmic
rays to diffuse into the magnetic compression. Our approach may be
contrasted with that of Kubo \& Shimazu, (2010), who use a
computational approach to investigate cosmic ray dynamics (thus
incorporating both the mirror effect and cross-field diffusion) in a
magnetic field configuration that comprises an idealized flux rope
CME. They do not consider the sheath region, and neither do they use
observations to guide their choice of magnetic field turbulence
levels.

To calculate the cross-field diffusion timescale, we proceed as
follows: considering the flux rope geometry of a near-Earth CME,
the magnetic field along the Sun-Earth ($\rm B_X$) represents the
longitudinal magnetic field. The fields $\rm B_y$ and $\rm B_z$
represent the perpendicular magnetic fields encountered by the
diffusing protons. In our discussion we consider only cross-field
diffusion; we therefore choose events that  exhibit good correlation
with the $\rm B_y$ and $\rm B_z$ magnetic field compressions and poor
correlation with compressions in $\rm B_x$. The events shortlisted
using these criteria are listed in Table\,\ref{T2}.

\begin{table*}
\centering
\caption{Events for which the FD profile correlates well only with the perpendicular component of the IMF enhancement. These events are a subset of the events listed in Table~\ref{T1}. The associated CMEs are also listed. The correlation lags are given in hours.}\label{T2}
\begin{tabular}{llccccccccc}\hline \hline
 Event  & CME & Time  & type & $\rm V_m $&\multicolumn{2}{c} {$\rm B_{total}$} & \multicolumn{2}{c} {$\rm B_{y}$} & \multicolumn{2}{c} {$\rm B_{z}$} \\ 
& (near Sun) & (UT) & & $\rm km \, s^{-1}$& Corr.(\%) & Lag (hrs) &  Corr.(\%) & Lag (hrs) &  Corr.(\%) & Lag (hrs)\\ \hline 
2001 Jan 13  & Jan 10 & 00:54 & Halo & 832 & 97.2 & -13 &  95.8 & -14 & 96.6 & -23 \\
2001 Apr 11  & Apr 10 & 05:30 & Halo & 2411 &  90.7 & -19  & 89.9  &  -17  &  91.3 &  -5 \\   
2001 May 27  & May 25 & 04:06 & 354 & 569 & 66.5 & 0 &  65.0 & -3 & 75.7 & -21 \\
2001 Aug 13  & Aug 11 & 04:30 & 313 & 548 & 51.8 & -7 & 97.5 & -7 & 70.0 & -5\\
2001 Sep 12  & Sep 11 & 14:54 & Halo & 791 & 79.2 & -25 & 26.7 & -30  & 86.2 & -1  \\
2001 Nov 24  & Nov 22 & 23:30 & Halo & 1437 & 85.3 & -21 & 41.0  & -31  & 77.1 & -14  \\ 
2001 Dec 14  & Dec 13 & 14:54 & Halo & 864 & 74.7 & -35 & 69.7 & -2 & 72.8 & -17\\ 
2002 Sep 07  & Sep 05 & 16:54 & Halo & 1748  & 77.1 & -18 & 49.6 & -24 & 87.4 & 3  \\
2002 Sep 30  & Sep 29 & 15:08 & 261 & 958 & 81.1 & -5 & 72.1 & 8 & 75.7 & -12\\
2002 Dec 22  & Dec 19 & 22:06 & Halo & 1092 &  73.4 & -15 & 43.4 & 0 & 84.7  & -12\\ 
2003 Jan 23  & Jan 22 & 05:06 & 338 & 875 & 70.9 & -21 & - & - & 75.4 & -28  \\
2003 Feb 16  & Feb 14 & 20:06 & 256 & 796 & - & - & - & - & 74.3 & -11  \\
2003 May 04  & May 02 & 12:26 & 222 & 595 & 83.4 & -8 & 84.7 & -10 & 80.7 & 0 \\
2003 Jul 25  & Jul 23 & 05:30 & 302 & 543  & 95.3 & -19 & 73.3 & -2 & 41.6  & 2  \\
2003 Dec 27  & Dec 25 & 09:06 & 257 & 178 & 86.1 & -35 & 21.7 & 5 & 87.2 & -3\\
2004 Aug 30  & Aug 29 & 02:30 & 274 & 1195 & - & - & -  & - & 92.4& 1\\
2004 Dec 05  & Dec 03 & 00:26 & Halo & 1216 & 85.3 & -12 & 89.4 & 8 & 58.4 & -13 \\
2004 Dec 12  & Dec 08 & 20:26 & Halo & 611  & 81.1 & -17 & 73.3 & -25& 78.9 & -13\\ \hline
\end{tabular} 
\end{table*}

\section{Cross-field diffusion coefficient ($\rm D_{\perp}$)} \label{dperp}  

The cross-field diffusion coefficient $\rm D_{\perp}$ governs the
diffusion of the ambient high-energy protons into the CME across the
magnetic fields that enclose it. The topic of cross-field diffusion
of charged particles across magnetic field lines in the presence of
turbulence is the subject of considerable research. Analytical treatments
include classical scattering theory (e.g., Giacalone \& Jokipii 1999,
and references therein) and non-linear guiding center theory
(Matthaeus et al. 2003; Shalchi, 2010) for cross-field diffusion.
Numerical treatments of cross-field diffusion of charged particle in
turbulent magnetic fields include Giacalone \& Jokipii (1999), Casse,
Lemoine, \& Pelletier (2002), Candia \& Roulet (2004), Tautz \& Shalchi
(2011), and Potgieter et al. (2014). We seek a concrete prescription for
$\rm D_{\perp}$ that can incorporate observationally determined
quantities. 

\subsection{$\rm D_{\perp}$: Candia \& Roulet (2004)} 
The $\rm D_{\perp}$ prescription we use is given by Candia \& Roulet
(2004), obtained from extensive Monte Carlo simulations of cosmic rays
propagating through tangled magnetic fields. Their results reproduce
the standard results of Giacalone \& Jokipii (1999) and Casse, Lemoine,
\& Pelletier (2002), and also extend the regime of validity to include
strong turbulence and high rigidities. The extent of cross-field
diffusion of protons depends on  the proton rigidity, which indicates
how tightly the proton is bound to the magnetic field, and the level of
magnetic field turbulence, which can contribute to field line transport. 

Candia \& Roulet (2004) give the following fit for the ``parallel"
diffusion coefficient $\rm D_{\parallel}$ (which is due to scattering
of the particles back and forth along the mean field, as the field is
subject to random turbulent fluctuations),

\begin{equation}
\textit D_{\parallel} \equiv c L_{max} \rho \frac {N_{\parallel}}{\sigma^2}\sqrt {\left( \frac {\rho}{{\rho}_{\parallel}} \right) ^{2(1-\gamma)} +\left( \frac {\rho}{{\rho}_{\parallel}} \right) ^ 2 } 
	\end{equation}

where c is the speed of light and the quantities $\rm N_{\parallel}$,
$\gamma$, and $\rm {\rho}_{\parallel}$ are constants specific to
different kinds of turbulence whose values are listed in Table\,1 of
Candia \& Roulet (2004). The parameter $\rm L_{max}$ is the maximum length scale of
turbulence; in our case we considered it as the size of the CME near the
Earth. The quantity $\rho$ is related to the rigidity of the proton
$\rm Rg$ as 

\begin{equation}
\rm \rho \, = \, \frac{r_L}{L_{max}} \, = \, \frac{Rg}{B_0 L_{max}}, 
\end{equation}

where $\rm r_L$ is the Larmor radius and $\rm B_0$ is the magnetic
field. The quantity $\sigma^2$ is the magnetic turbulence level,
which is defined as in Eq. \ref{dysig}.

The cross-field diffusion coefficient ($\rm D_{\perp}$) is related to
the parallel one ($\rm D_{\parallel}$ ) by,

\begin{equation}
\nonumber\rm 
	\frac {D_{\perp}}{D_{\parallel}} \equiv \cases{ N_{\perp} {\left(\sigma^2 \right)}^{a_{\perp}},  &  $( \rho \le 0.2)$ \cr
\noalign{\medskip}
N_{\perp} { \left( \sigma^2 \right)}^{a_{\perp}} \left( \frac {\rho}{0.2} \right)^{-2},  & $( \rho \ge 0.2)$} 
\end{equation}

The quantities $\rm N_{\perp}$ and $\rm a_{\perp}$ are constants
specific to different kinds of turbulent spectra, and are given
in Table\,1 of Candia \& Roulet (2004). { We note that the exponent characterizing the IMF turbulence may vary through the magnetic field compression associated with FD (Alania \& Wawrzynczak, 2012). Although the treatment of Candia \& Roulet (2004), which we use, cannot accommodate arbitrary turbulent spectrum indices, it can address the Kolmogorov ($\gamma$ = 5/3), Kraichnan ($\gamma$ = 3/2), and Bykov–-Toptygin ($\gamma$ = 2) spectra. We therefore quote results for all three turbulence spectra.}

\section{The IP B field compression-FD lag: how many diffusion lengths?}\label{NdL} 
We have shown that the FD profile is often very similar to that of
the IMF compression, and lags behind it by a few hours. This observed lag is poorly correlated with the FD magnitude and the CME speeds (both near the Sun and near the Earth). We interpret the
observed time lag between the IMF and the FD profiles as the time
taken by the protons to diffuse through the magnetic field compression
via cross-field diffusion. The time taken for a single diffusion random
walk of a high-energy proton into the magnetic structure of CME is
given by

\begin{equation}
t_{diff} = \frac{D_{\perp}}{c V_{sw}} ,
\end{equation}

where c is the speed of light (which is the typical propagation speed for
the highly relativistic galactic cosmic rays we are concerned with)
and $\rm V_{sw}$ is the solar wind velocity upstream of the CME. 

{
When using
the $\rm D_{\perp}$ from Candia \& Roulet (2004), we use two different
methods for computing the turbulence level $\sigma$. In the first one,  we
calculate $\sigma$ as a function of time using the one-minute averaged
IMF data, as described in Eq\,(\ref{dysig}). In the second, we assume a constant value of 15\% for $\sigma$, which is typically expected to be the maximum level of turbulence in quiet solar wind (Spangler, 2002). We used both these methods to calculate $\rm t_{diff}$ for Kolmogorov, Kraichnan, and Bykov-Toptygin turbulent spectra}.
Using these values of $\rm t_{diff}$, we estimated the number of diffusion
lengths required to account for the observed time lag between the FD
profile and the IMF profile using

\begin{equation}
{\rm No. \,  of \, Diffusions \, = \frac{Lag}{t_{diff}} }\label{noDs}
\end{equation}

The results for the number of diffusion times needed to account for the
observed lag between the IMF enhancement and the FD profile are shown
in Table\,\ref{T3}. These numbers are calculated using the peak value
of the IMF profile. It is evident that the observed lags can be
accounted for by a few tens to a few hundred diffusion times. There
are two exceptional events on 2001 December 14 and 2003 December 27,
where the number of diffusions are $\sim$1000 using the time-varying
$\sigma$ prescription. There are three events in this list that have
no correlation lag between the IMF profile and FD profile. The FD on
2001 May 27 correlates well with $\rm B_{total}$, the FD on 2002
December 22 correlates well with  $\rm B_y$, and the FD on 2003 May
04 correlates with the $\rm B_z$ with no correlation lag.

\section{Summary}
We studied all FD events observed by the GRAPES-3 muon telescope
during the years 2001--2004 satisfying the broad criteria listed in \S~\ref{SL1}. For a sample of especially well-observed events, we find that the magnetic field compression
responsible for the FD as well as the turbulence level gets enhanced
in the shock-sheath region. For these events, details regarding shock timing,
magnetic cloud start and end timings along with the FD onset time for
different directions are given in Table\,\ref{tabtim}.

We find that the FD profile looks remarkably similar to that of the
corresponding IMF compression and lags behind it by few hours (Table~\ref{T1}). { Since we want to focus on cross-field diffusion, we
 selected the FD events whose profiles correlate well with the
enhancements in the perpendicular magnetic fields ($\rm B_y$,
$\rm B_z$) and not with the radial magnetic field ($\rm B_x$); these events are listed in Table~\ref{T2}. We
have calculated the number of diffusions using Eq\,\ref{noDs} for
14.3 GV and 24.0 GV protons. The number of diffusions corresponding to the observed lag
for the selected events are listed in Table\,\ref{T3}.} For most events
we find that the observed time lag corresponds to a few tens to a few hundred diffusions.

\section{Conclusion}
The results of Arunbabu et al. (2013) show that FDs are due to
cumulative diffusion of galactic cosmic ray protons into the CME as
it propagates from the Sun to the Earth. However, the precise nature
of the diffusive barrier was left unspecified, and the diffusion was
assumed to occur across an idealized thin boundary that presumably
had to do with the turbulent sheath region. The results from this
work clearly show that the magnetic field enhancement responsible
for the FD comprises the sheath region. The FD profile looks like a
lagged (and inverted) copy of the magnetic field enhancement
(Table\,\ref{corrT}). The FD lags behind the magnetic field enhancement by
a few hours (Tables\,\ref{T1} and \ref{T2}). We have quantitatively
shown that the time lag between the FD and the magnetic field
enhancement can be accounted for by cross-field diffusion through
the turbulent sheath region (Table\,\ref{T3}). {  This work establishes
i) the importance of the turbulent sheath region between the shock and
ICME; we show that the magnetic field enhancement responsible for
the FD comprises the shock-sheath, and the magnetic turbulence level
is also enhanced in this region {\bf (\S~\ref{ipmc})} and  
ii) the viability of cross-field diffusion through the turbulent CME sheath
as the primary mechanism for FDs {\bf (\S~\ref{NdL})}. }
 

\begin{acknowledgements}
K. P. Arunbabu acknowledges support from a Ph.D. studentship at IISER
Pune. P. Subramanian acknowledges partial support via the CAWSES-II
program administered by the Indian Space Research Organization and via
a grant from the Asian Office of Aerospace Research and Development,
Tokyo. We thank D. B. Arjunan, A. Jain, the late S. Karthikeyan, K.
Manjunath, S. Murugapandian, S. D. Morris, B. Rajesh, B. S. Rao, C.
Ravindran, and R. Sureshkumar for their help in the testing,
installation, and operation of the proportional counters and the
associated electronics and during data acquisition. We thank G. P.
Francis, I. M. Haroon, V. Jeyakumar, and K. Ramadass for their help
in the fabrication, assembly, and installation of various mechanical
components and detectors. We are thankful to the Tibet neutron
monitor groups for making the data available on the internet. We thank the anonymous referee for a thorough and helpful review.
\end{acknowledgements}

\appendix

\section{Additional figures}
\begin{figure*}[h]
\centering
\includegraphics[width = 0.8\textwidth]{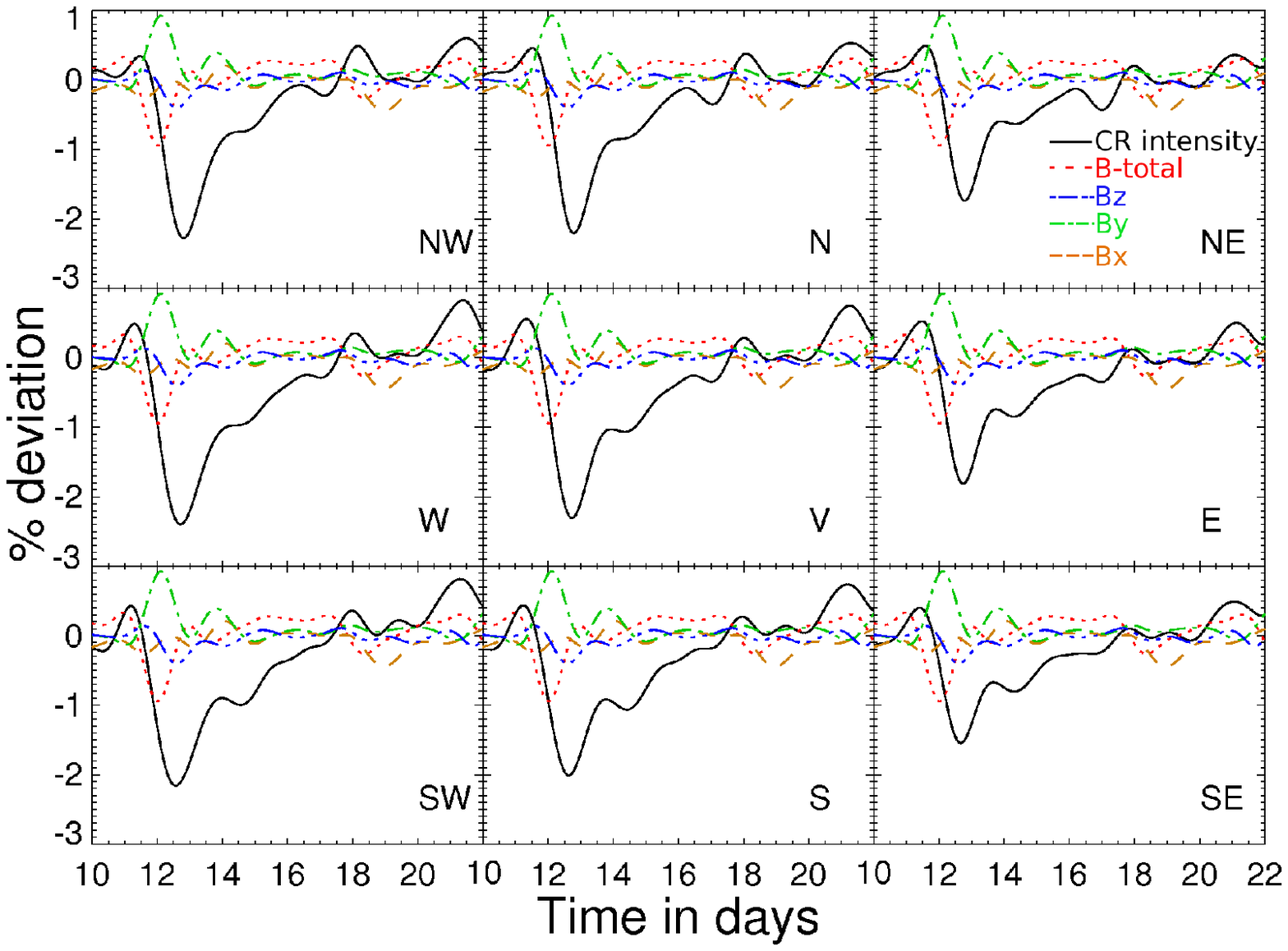}
\caption{The FD event of { 11 April 2001 } and the magnetic field for
         nine directions in GRAPES-3 muon telescope. The linestyles are the same as used in Figure \ref{FDi} }\label{FDiap11}
\end{figure*}

\begin{figure*}[h]
\centering
\includegraphics[width = 0.8\textwidth]{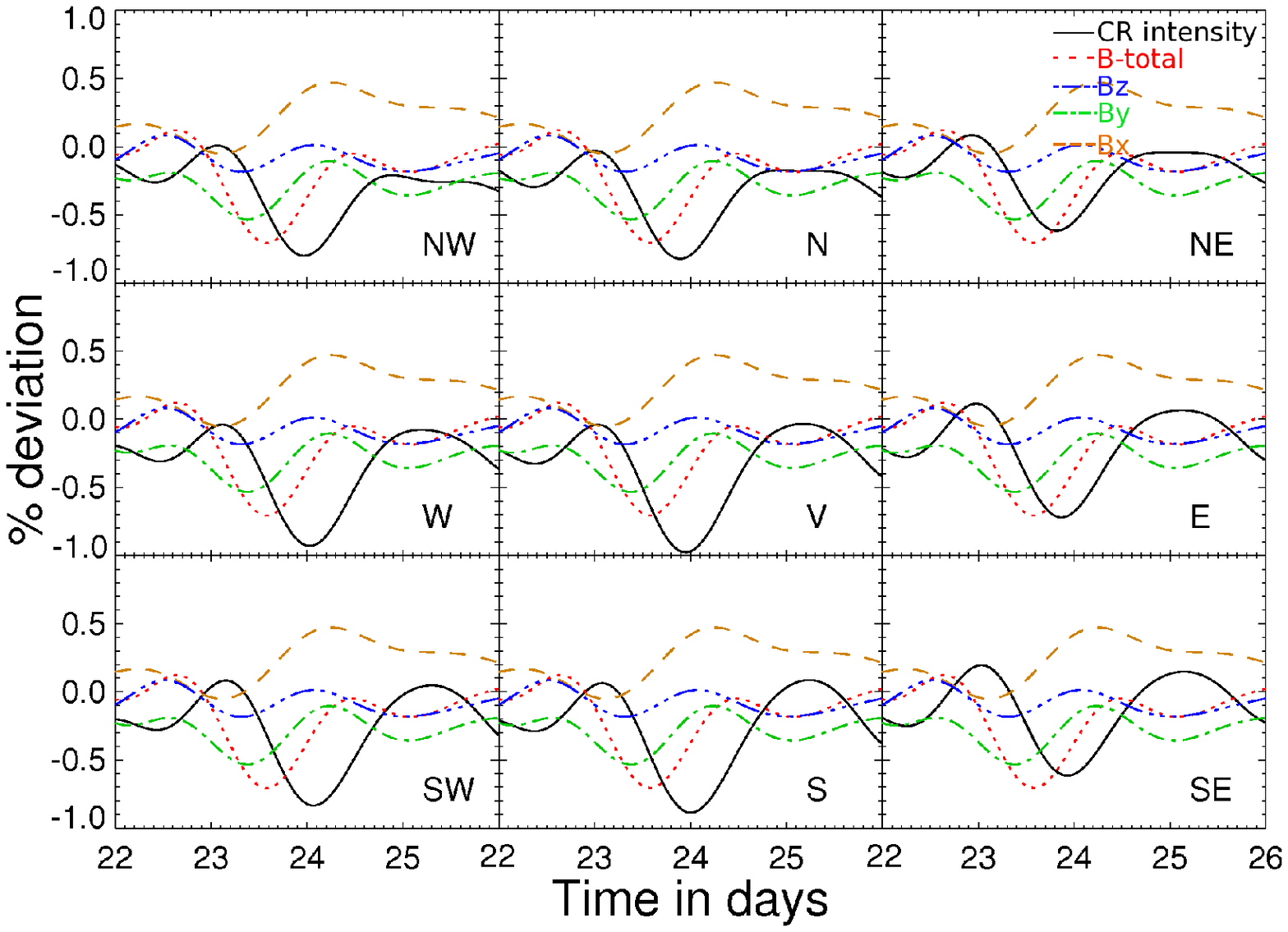}
\caption{The FD event of {23 May 2002 } and the magnetic field for
         nine directions in GRAPES-3 muon telescope. The linestyles are the same as used in Figure \ref{FDi} }\label{FDimy23}
\end{figure*}

\begin{figure*}[h]
\centering
\includegraphics[width =0.8\textwidth]{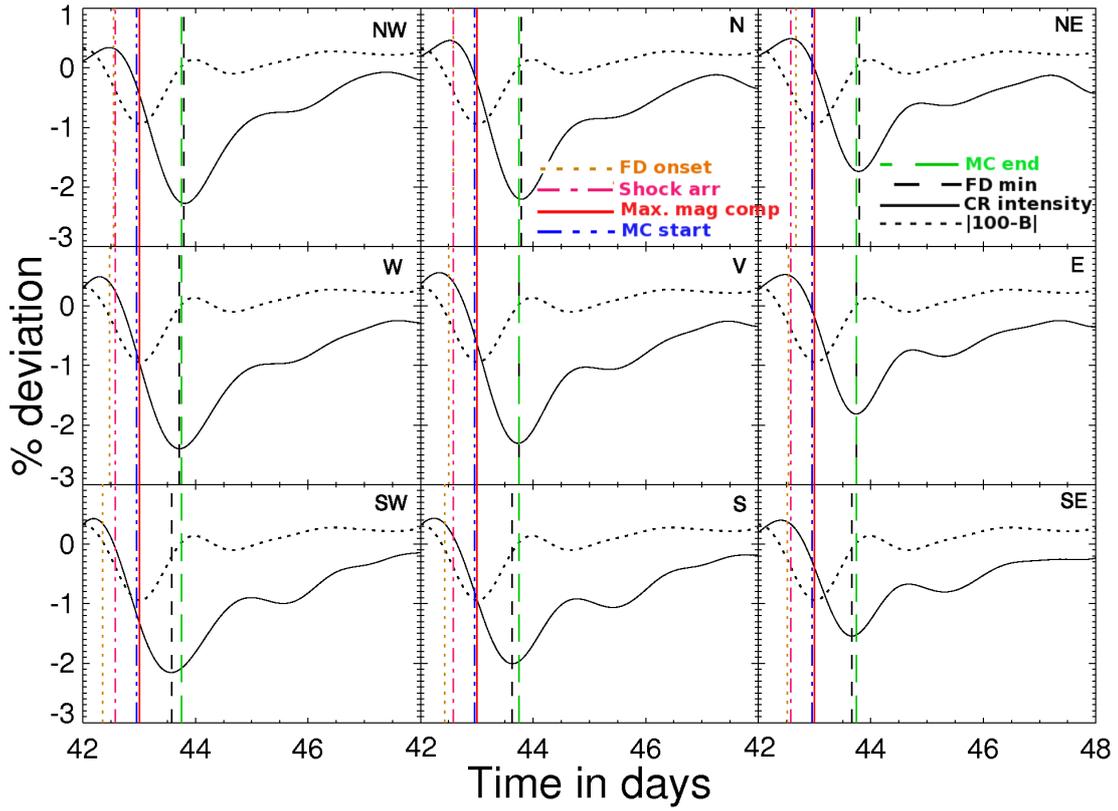}
\caption{FD event on 11 April 2001. The linestyles are the same as used in figure \ref{timi} }\label{timiapr11}
\end{figure*}

\begin{figure*}[h]
\centering
\includegraphics[width =0.8\textwidth]{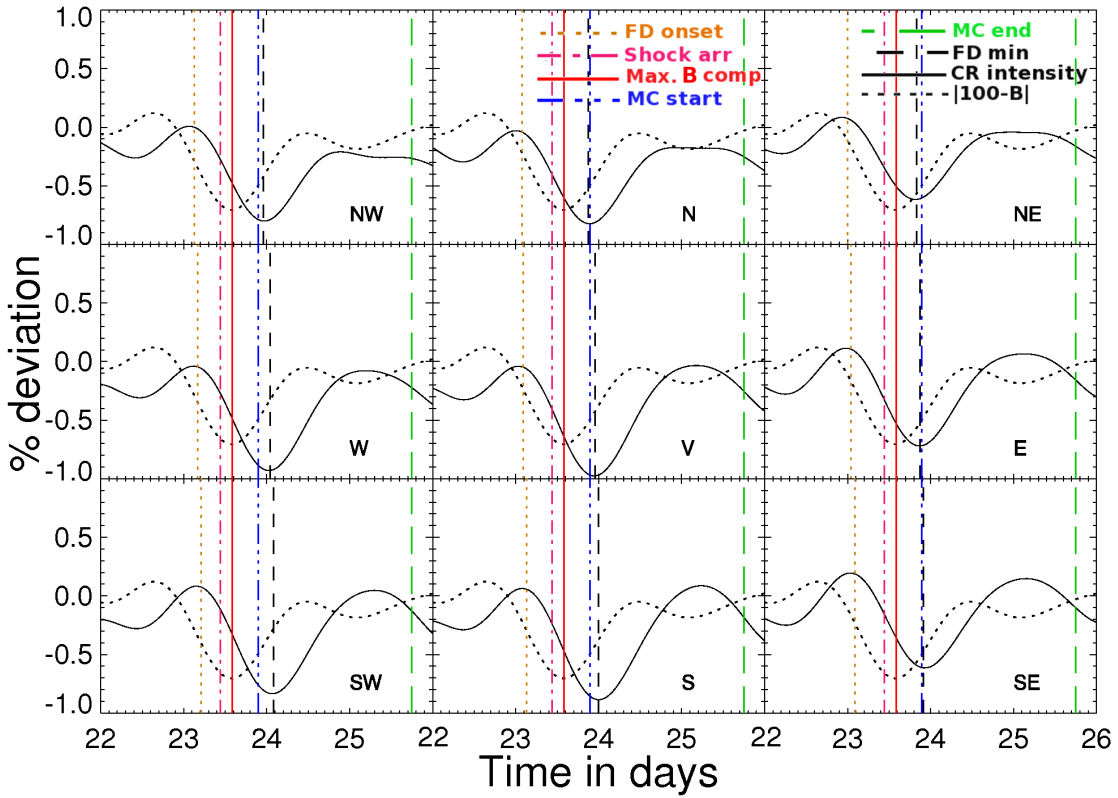}
\caption{FD event on 23 May 2002. The linestyles are the same as used in figure \ref{timi} }\label{timimay23}
\end{figure*}

\begin{figure*}[h] 
\centering
\includegraphics[width =0.7\textwidth]{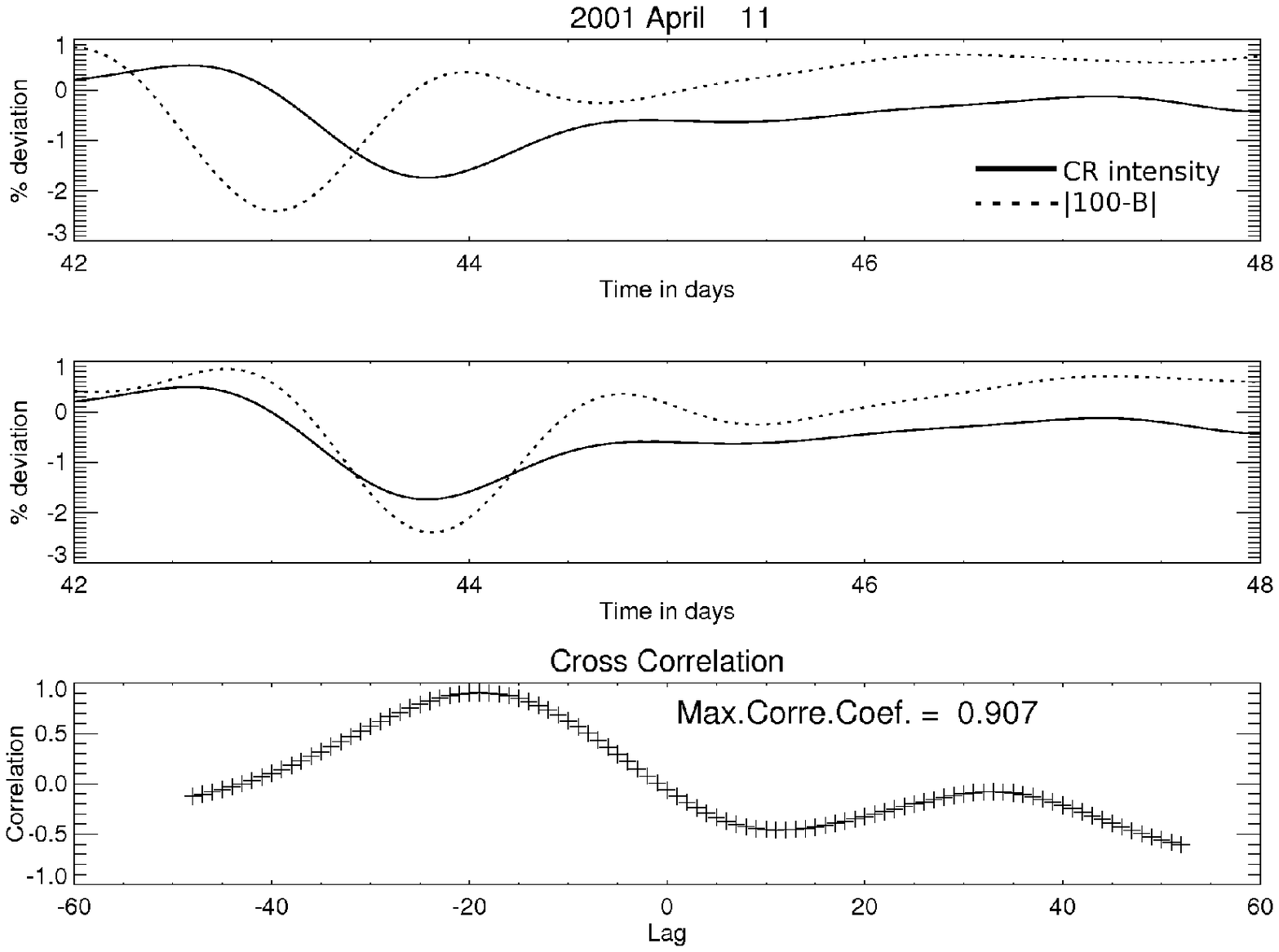}
\caption{Cross correlation of the cosmic ray flux with $\rm B_{total}$.
         The top panel shows the percentage deviation of cosmic ray flux
         using solid black lines and the magnetic field using dotted
         black lines (scaled to fit in the frame). The middle panel
         shows the same with magnetic field shifted to the right
         corresponding to the time lag and the bottom panel shows the
         correlation coefficient for different lags.}\label{11apr} 
\end{figure*}

\begin{figure*}[h]
\centering
\includegraphics[width =0.7\textwidth]{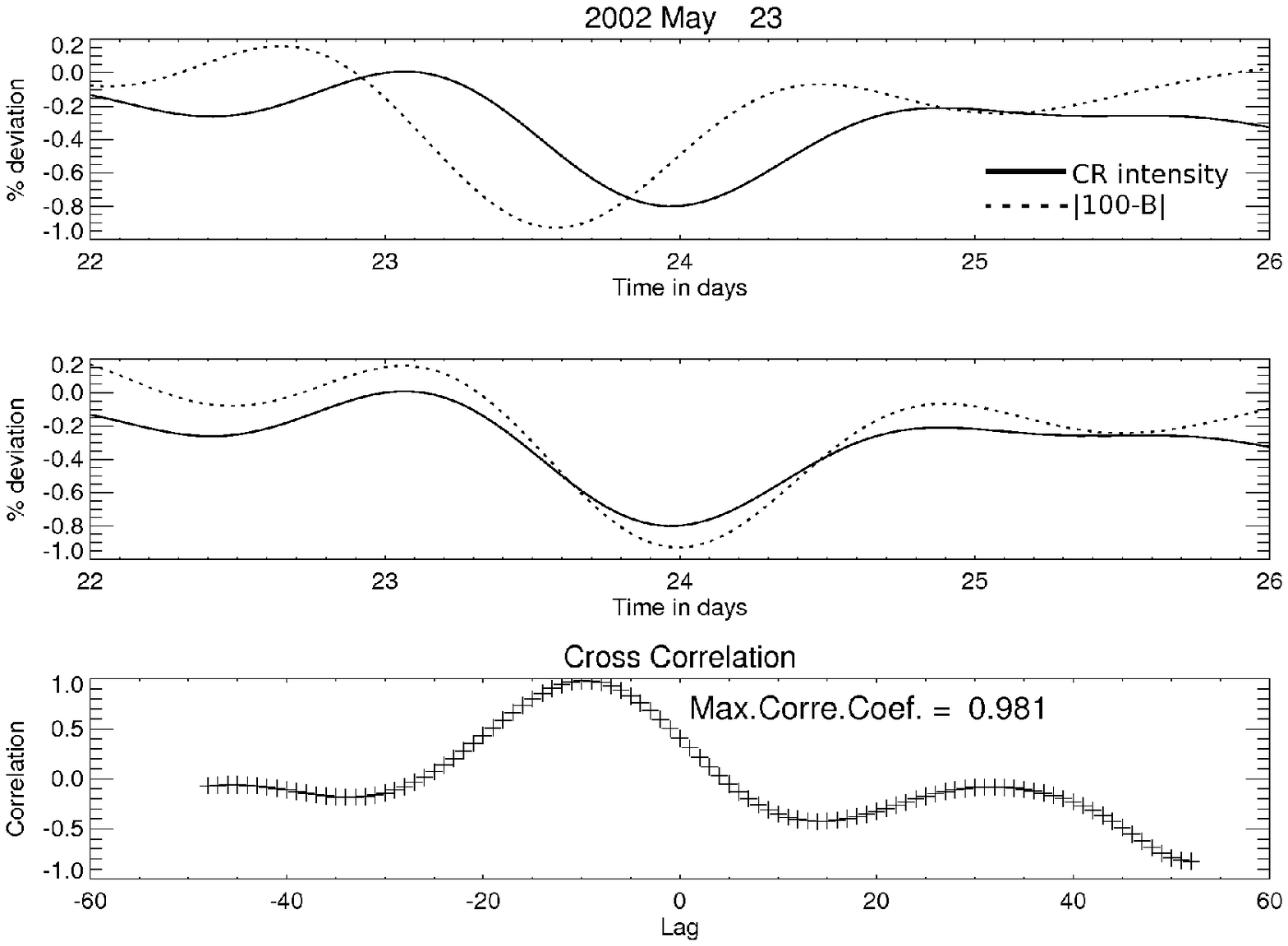}
\caption{Cross correlation of the cosmic ray flux with $\rm B_{total}$.
         The top panel shows the percentage deviation of cosmic ray flux
         using solid black lines and the magnetic field using dotted
         black lines (scaled to fit in the frame). The middle panel
         shows the same with magnetic field shifted to the right
         corresponding to the time lag and the bottom panel shows the
         correlation coefficient for different lags.}\label{23may} 
\end{figure*}

\section{Additional Tables}

\begin{table*}
\centering
\caption{ List of FD events for which the correlation coefficient between the profiles of the FD and the IMF enhancement $\geq$\,70\%. The `-'
         entries denote events that have low correlation values for
         lags between -36 and 12 hours} \label{T1}
\begin{tabular}{lcccccccc} \hline \hline
Event 		& \multicolumn{8}{c} {Correlation (\%)}\\ 
      		& \multicolumn{2}{c}{$\rm B_{total} $} & \multicolumn{2}{c}{$\rm B_{x} $ }& \multicolumn{2}{c}{$\rm B_{y} $ }& \multicolumn{2}{c}{$\rm B_{z} $} \\ 
      		& coeff. & err  & coeff. & err  & coeff. & err  & coeff. & err\\  \hline \hline
13 Jan 2001 	& 97.2 	& 2.4 	& - 	&  -  	& 95.8 	& 2.9 	& 96.6 	& 2.6 \\
26 Mar 2001 	& 70.3 	& 6.5 	& - 	& - 	& - 	& - 	& 39.1 	& 8.3 \\
4 Apr 2001 	& 92.8 	& 3.1 	& 97.3 	& 1.9 	& 77.2 	& 5.3 	& 63.4 	& 6.4 \\
7 Apr 2001 	& 94.3 	& 3.0	& 92.7 	& 3.4	& 71.9 	& 6.2	& 54.2  & 7.7 \\
11 Apr 2001 	& 90.7  & 3.5	& - & -	& 89.9  & 3.7 	& 91.3  & 3.4	 \\
27 May 2001 	& 66.5 	& 7.7	& 27.8 	& 9.8	& 65.0  & 7.7	& 75.7  & 6.7	\\
 1 Jun 2001 	& 77.1	& 4.9	& 70.5 	& 5.4	& 52.3 	& 6.6	& 54.5 	& 6.5	\\
13 Aug 2001 	& 51.8 	& 7.2	& - 	& -	& 97.5 	& 1.8	& 70.0 	& 5.9	\\
17 Aug 2001 	& 84.6 	& 3.8	&  - 	& -	& 31.2 	& 6.8	& 58.0 	& 5.9	\\
6 Sep 2001 	& 68.8 	& 6.1	& 87.0 	& 4.1	& 64.7 	& 6.3	& 45.1	& 7.5	\\
 12 Sep 2001  	& 79.2 	& 4.2	& 56.2 	& 5.6	& 26.7 	& 6.6	& 86.2 	& 3.5	\\
29 Sep 2001 	& 70.3 	& 4.0	& - 	& -	& 58.1 	& 4.6	& - 	& -	\\
5 Nov 2001 	& 88.3 	& 3.0	& 64.6 	& 4.9	& 34.8 	& 6.1	& -  	& -	\\
24 Nov 2001 	& 85.3 	& 3.4	& 32.4 	& 6.1	& 41.0  & 5.9	& 77.1 	& 4.1	\\
14 Dec 2001 	& 74.7 	& 3.5	& 42.7 	& 4.8	& 69.7 	& 3.8	& 72.8 	& 3.6	\\
23 May 2002 	& 98.1  & 2.0	& 79.4 	& 6.2	& 75.9 	& 6.7	& 60.0  & 8.2	\\
 7 Sep 2002 	& 77.1 	& 4.6	& - 	& -	& 49.6 	& 6.3	& 87.4 	& 3.5	\\
23 Sep 2002 	& 60.4 	& 5.8	& 87.9 	& 3.5	&  41.4 & 6.6	& 93.1 	& 2.6	\\
30 Sep 2002 	& 81.1  & 5.3	& 58.7 	& 7.4	& 72.1 	& 6.3	&  75.7 & 6.0	\\
22 Dec 2002 	& 73.4 	& 4.9	& -  	& -	& 43.4 	& 6.5	& 84.7 	& 3.8	\\
9 Jan 2003 	& 90.1 	& 3.3	& 68.1 	& 6.7	& - 	& -	& 56.2 	& 6.4	\\
 23 Jan 2003 	& 70.9 	& 6.4	& - 	& -	& - 	& -	& 75.4 	& 6.0	\\
30 Jan 2003 	& 94.8 	& 3.7	& 84.4 	& 6.4	& 42.7 	& 10.7	& 95.7 	& 3.4	\\
16 Feb 2003 	& - 	& -	& 31.3 	& 5.3	& - 	& -	& 74.3	& 3.7	\\
26 Mar 2003 	& 77.1 	& 3.8	& 64.8 	& 4.5	& - 	& -	& - 	& -	\\
4 May 2003 	& 83.4 	& 5.6	& 32.8  & 9.7 	& 84.7 	& 5.4	& 80.7  & 6.1	\\
18 May 2003 	& 86.5 	& 3.6	& -  	& -	& - 	& -	& -  	& -	\\
25 Jul 2003 	& 95.3 	& 2.8	& 47.9 	& 8.0	& 73.3 	& 6.2	&  41.6 & 8.3	\\
16 Aug 2003 	& 71.6	& 6.4	& 49.6 	& 8.0	& 45.5 	& 8.2	& 57.7 	& 7.5	\\
21 Oct 2003 	& 80.1 	& 6.1	& 92.0 	& 4.0	& 70.1 	& 7.3	& 93.5  & 3.6	\\
27 Dec 2003 	& 86.1 	& 4.2	& - 	& -	& 21.7 	& 8.1	& 87.2 	& 4.1	\\
 21 Jan 2004 	& 77.9 	& 4.0	& 78.2 	& 4.0	& - 	& -	& -  	& -	\\
29 May 2004 	& 53.1  & 6.6	& 90.2 	& 3.3	& - 	& -	& 86.9 	& 3.8	\\
26 Jul 2004 	& 86.5 	& 4.2	& 73.3 	& 5.7	& 85.5 	& 4.3	& 94.8 	& 5.6	\\
30 Aug 2004 	& -  	& -	& - 	& -	& -  	& -	& 92.4 	& 2.9	\\
5 Dec 2004 	& 85.3 	& 3.1	& - 	& 	& 89.4 	& 2.6	& 58.4 	& 4.8	\\
12 Dec 2004 	& 81.1 	& 4.2	& 61.6  & 5.7	& 73.3 	& 4.9	& 78.9 	& 4.4	\\ \hline \hline
\end{tabular}
\end{table*}

\begin{landscape}
\begin{table}
{
\caption{\label{T3} Number of diffusions required for the observed
         lag in FD events using three different turbulent spectrums and two different methods described
         in \S\ref{NdL}. $\rm 1^{st}$ stands for the method using
         dynamic $\sigma$, $\rm 2^{nd}$ stands for the method using
         a constant $\sigma$.}
\begin{tabular}{|l|c|c|c|c|c|c|c|c|c|c|c|c|c|c|c|c|c|c|c|c|c|c|c|}

\hline \hline
     & $\sigma$    & Rg &  \multicolumn{7}{|c|} {$\rm B_{total}$}    &  \multicolumn{7}{|c|} {$\rm B_{y}$} &  \multicolumn{7}{|c|} {$\rm B_{z}$} \\ \cline{4-24}
 Event & &  & Lag & \multicolumn{6}{|c|} {No: of diffusions} & Lag & \multicolumn{6}{|c|} {No: of diffusions} & Lag & \multicolumn{6}{|c|} {No: of diffusions}\\ \cline{5-10} \cline{12-17} \cline{19-24}  
      &(\%) & (GV) & (hrs) &  \multicolumn{2}{|c|} {Kolmo.} &  \multicolumn{2}{|c|} {Kraich.} &  \multicolumn{2}{|c|} {Bykov.} & (hrs) &  \multicolumn{2}{|c|} {Kolmo.} &  \multicolumn{2}{|c|} {Kraich.} &  \multicolumn{2}{|c|} {Bykov.} & (hrs) &   \multicolumn{2}{|c|} {Kolmo.} &  \multicolumn{2}{|c|} {Kraich.} &  \multicolumn{2}{|c|} {Bykov.}  \\ \cline{5-10} \cline{12-17} \cline{19-24} 
      & &     & &  $\rm 1^{st}$ &$\rm 2^{nd}$  &  $\rm 1^{st}$ &$\rm 2^{nd}$  &  $\rm 1^{st}$ &$\rm 2^{nd}$  & & $\rm 1^{st}$ &$\rm 2^{nd}$  &  $\rm 1^{st}$ &$\rm 2^{nd}$ &  $\rm 1^{st}$ &$\rm 2^{nd}$  & &  $\rm 1^{st}$ &$\rm 2^{nd}$  &  $\rm 1^{st}$ &$\rm 2^{nd}$  &  $\rm 1^{st}$ &$\rm 2^{nd}$   \\ 
\hline  \hline

2001  Jan 13  &8-15 &14.3 &  -13 & 384 & 173 & 462 & 210 & 644 & 265 & -14 & 196 & 158 &  232 & 186 & 314 & 251 & -23 &  97 & 112 & 115 & 133 & 133 & 155\\ 
              & &24.0 &  -13 & 245 & 121 & 292 & 140 & 395 & 188 &  -14 & 117 &  95 &  137 & 110  & 172 & 137 & -23 &  38 &  44 & 47 & 55 & 49 & 57\\ \hline
2001  Apr 11  &10-23 &14.3 &  -19 & 656 & 340 & 895 & 482 & 907 & 396 &  -17 &  326 &  292 &  425 & 394 & 453 & 372 & -5 &  41 &  67 & 47 & 81 & 64 & 102 \\ 
              & &24.0 &  -19 & 533 & 283 & 670 & 367 & 844 & 390 & -17 &   258 &   239 &  312 & 296 & 409 & 356 & -5 &  25 &  47 & 28 & 54 & 37 & 73 \\  \hline 
2001  May 27  &8-13 &14.3 &   0 &  0 &  0 &  0 & 0 & 0 & 0 &  -3 &  44 &  38 &  53 & 46 &  71 & 59 & -21 & 111 & 177 & 128 & 205 & 165 & 270\\ 
              & &24.0 &  0 &  0 &  0 &  0 & 0 & 0 & 0 &     -3 &  29 &  25 & 34 & 29 & 45 & 38 &      -21 &  53 &  85 & 63 & 101 & 70 & 114  \\ \hline 
2001  Aug 13  &10-20 &14.3 &   -7 & 216 & 106 & 279 & 135 & 326 & 150 &     -7 &  76 &  86 & 91 & 102 &   120 & 135 &    -5 &  26 &  58 & 30 & 69   & 40 & 92 \\
              & &24.0 &   -7 & 165 &  82 & 199 & 97 &  273 & 130 &     -7 &  49 &  55 & 57 & 64 &  73 & 83 &     -5 &  15 &  35 & 17 & 41  & 21 & 52  \\ \hline 
2001  Sep 12  &10-20 &14.3 &  -25 & 838 & 354 & 1046 & 441 &  1357 & 526 &  - &   - &   - &   - & - & - & - &    -1 &  17 &  13 & 20 & 16 &  26 & 20\\
              & &24.0 &  -25 & 583 & 263 & 695 & 307 & 953 & 415 & - &   - &   - &   - & - & - & - &    -1 &  11 &   9 & 13 & 11 &  18 & 14  \\ \hline 
2001  Nov 24  &20-40 &14.3 &  -21 & 822 & 405 & 1119 & 578 &  1137 & 462 &   -31 & 173 & 429 & 205 & 529 &   241 & 646 & -14 & 260 & 226 & 331 & 296 &  381 & 301\\
              & &24.0 &  -21 & 666 & 338 & 834 & 442 & 1067 & 456 &    -31 & 112 & 311 & 126 & 363 & 165 & 490 &     -14 & 199 & 181 & 235 & 219 & 319 & 281   \\  \hline 
2001  Dec 14  &5-12 &14.3 &  -35 &1346 & 604 & 1644 & 817 & 2283 & 766 &    -2 &  56 &  28 &  67 & 35 &  92 & 42 & -17 & 255 & 240 &  305 & 298  & 396 & 357\\
              & &24.0 &  -35 & 872 & 494 & 1066 & 614 &  1440 & 736 &  -2 &  28 &  21 & 34 & 25 &  41 & 33 &    -17 & 119 & 177 & 145 & 206  & 165 & 279  \\  \hline 
2002  Sep 07  &5-20 &14.3 &  -18 & 776 & 298 &  983 & 396 & 1255 & 393 &  -24 & 312 & 305 & 367 & 368 &  499 & 476 &    3 &  24 &  37 &  28 & 45 &  38 & 59\\
              & &24.0 &  -18 & 555 & 241 & 667 & 295 & 928 & 369 &   -24 & 172 & 205 & 204 & 238 & 245 & 313 &     3 &  14 &  25 &  16 & 29 & 20 & 37  \\ \hline 
2002  Sep 30  &10-20 &14.3 &   -5 & 207 &  93 &  294 & 131  & 362 & 210 & 8 & 130 & 128 &  160 & 167 &  198 & 174 &  -12 & 198 & 185 & 243 & 238  & 309 & 259 \\
              & &24.0 &   -5 & 171 &  77 & 224 & 99 &  254 & 108 &    8 &  93 & 102 &  109 & 123 &  148 & 159 &  -12 & 138 & 145 & 162 & 173 &  217 & 229 \\ \hline 
2002  Dec 22  &5-13 &14.3 & -15 & 481 & 244 & 617 & 321 & 779 & 327 &  0 &  0 &  0 &  0 & 0 & 0 & 0 &    -12 &  66 & 126 & 75 & 147 &  100 & 199 \\
	      & &24.0 &  -15 & 363 & 196 & 436 & 238 &  603 & 303 &  0 &  0 &  0 &  0 & 0 & 0 & 0 &    -12 &  37 &  70 & 42 & 82 &  50 & 99 \\ \hline 
2003  Jan 23  & 8-18&14.3 &  -21 & 509 & 270 &  624 & 327 & 823 & 420 & - &   - &   - & - & - & - & - &     -28 &  49 & 170 & 56 & 200 &  66 & 243  \\
              & &24.0 &  -21 & 341 & 183 &  403 & 213 & 540 & 281 & - &   - &   - & - & - & - & - &      -28 &  21 &  71 & 24 & 87 &  25 & 93  \\ \hline 
2003  Feb 16  & 5-17 &14.3 &   - &   - &   - & - & - & - & - &     - &   - &   - & - & - & - & - &     \-11 &  55 & 100 & 64 & 117 & 76 & 155   \\
              & &24.0 &  - &   - &   - & - & - & - & - &    - &   - &   - & - & - & - & - &      -11 &  22 &  50 & 27 & 60 &  29 & 69\\ \hline 
2003  May 04  & 10-40&14.3 &   -8 & 252 & 127 &  334 & 165 & 374 & 174 &  -10 &  97 & 125 &  115 & 149 & 150 & 195 &  0 &  0 &  0 &  0 & 0 & 0 & 0  \\
              & &24.0 &   -8 & 200 & 101 &   245 & 121 & 325 & 157 &   -10 &  64 &  82 &   73 & 95 &  95 & 124 &  0 &  0 &  0 &  0 & 0 & 0 & 0  \\ \hline 
2003  Jul 25  & 8-15 &14.3 &  -19 & 565 & 347 &  732 & 483 & 861 & 417 &    -2 &  32 &  31 &  39 & 40 &  50 & 43 &    2 &   9 &  22 & 10 & 26 &  13 & 35  \\
              & &24.0 &  -19 & 434 & 288 & 554 & 367 & 704 & 408 &      -2 &  22 &  24 &  26 & 29 & 34 & 38 &  2 &   5 &  13 & 6 & 15 &  7 & 19 \\ \hline 
2003  Dec 27  & 4-12&14.3 &  -35 &1210 & 506 &  1546 & 634 & 1881 & 742 &    5 &  41 &  52 & 48 & 61 & 64 & 82 &     -3 &  10 &  27 & 11 & 31 &  16 & 19 \\
              & &24.0 &  -35 & 900 & 380 &  1079 & 446 &  1497 & 603 &  5 &  23 &  29 &  27 & 34 &  32 & 41 &   -3 &   5 &  13 & 6 & 16 & 6 & 18   \\ \hline 
2004  Aug 30  &8-20 &14.3 &  - &   - &   - & - & - & - & - &    - &   - &   - & - & - & - & - &     1 &  10 &  12 & 12 & 15 & 16 & 19  \\
              & &24.0 &  - &   - &   - & - & - & - & - &    - &   - &   - & - & - & - & - &     1 &   7 &   8 & 8 & 9 & 10 & 12  \\  \hline 
2004  Dec 05  &10-23 &14.3 &  -12 & 326 & 209 & 435 & 285 &  530 & 363 &   8 & 107 & 112 & 128 & 140 &  168 & 198 &   -13 & 117 & 193 & 146 & 244 & 166 & 278\\
              & &24.0 &  -12 & 260 & 172 & 327 & 215 & 403 & 254 &    8 &  71 &  83 &  82 & 97 &  107 & 131 &   -13 &  89 & 148 & 104 & 174 & 138 & 235 \\ \hline 
2004  Dec 12  &5-20 &14.3 &  -17 & 595 & 266 & 786 & 344 &  863 & 368 & -25 & 243 & 322 &  290 & 389 &  373 & 500 & -13 & 138 & 163 & 165 & 196 & 214 & 255 \\
              & &24.0 &  -17 & 469 & 210 &  573 & 251 & 771 & 330 & -25 & 163 & 218 &  187 & 253 & 245 & 335 &  -13 &  91 & 108 &  106 & 126 & 137 & 164 \\ \hline \hline 
\end{tabular} }
\end{table}
\end{landscape}

\end{document}